\documentclass[12pt]{article}
\usepackage{amssymb,amsmath}
\usepackage{amsthm}
\usepackage{xcolor}
\usepackage{url}
\newtheorem{theorem}{Theorem}

\theoremstyle{definition}

\newcommand{\beq}{\begin{equation}}
\newcommand{\eeq}{\end{equation}}
\def\d{\partial}
\def\f{\frac}

\begin{document}

\title{Bi-Hamiltonian structures of KdV type} \author{P. Lorenzoni ${}^{*}$,
  A. Savoldi ${}^{**}$ and R. Vitolo ${}^{***}$} \date{}
\maketitle
\vspace{-7mm}
\begin{center}
  ${}^{*}$ Dipartimento di Matematica e Applicazioni, University of Milano-Bicocca\\
  via Roberto Cozzi 53 I-20125 Milano, Italy\\
  ${}^{**}$ Department of Mathematical Sciences, Loughborough University \\
  Leicestershire LE11 3TU, Loughborough, United Kingdom \\
  ${}^{***}$ Dipartimento di Matematica e Fisica ``E. De Giorgi'', Universit\`a
  del Salento\\
  and Sezione INFN di Lecce\\
  via per Arnesano, 73100 Lecce, Italy
  \\
  e-mails:
  \texttt{paolo.lorenzoni@unimib.it}\\
  \texttt{A.Savoldi@lboro.ac.uk}\\
  \texttt{raffaele.vitolo@unisalento.it}
\end{center}

\bigskip


\begin{abstract}
  Combining an old idea of Olver and Rosenau with the classification of second
  and third order homogeneous Hamiltonian operators we classify compatible
  trios of two-component homogeneous Hamiltonian operators. The trios yield
  pairs of compatible bi-Hamiltonian operators whose structure is a direct
  generalization of the bi-Hamiltonian pair of the KdV equation. The
  bi-Hamiltonian pairs give rise to multi-parametric families of bi-Hamiltonian
  systems. We recover known examples and we find new integrable systems whose
  central invariants are non-zero; this shows that new examples are not
  Miura-trivial.
\end{abstract}

\section{Introduction}

Many integrable systems admit a bi-Hamiltonian structure. This means that these
systems can be written as Hamiltonian differential equations by means of two
compatible Hamiltonian operators $P$ and $Q$.

It was observed in \cite{OR} that in many examples the bi-Hamiltonian
structures are, in fact, defined by a compatible trio of Hamiltonian
operators. Usually $P$ is a first-order Hamiltonian operator and $Q$ is the sum
of a first-order Hamiltonian operator and a higher-order Hamiltonian operator,
and the three operators are mutually compatible. All these operators are
homogeneous in the sense of Dubrovin and Novikov \cite{DN,DN2}.

For instance, in the scalar case, one has the trio
\begin{equation}
  P=P_1=\d_x,\qquad Q=Q_1+R_3,\quad Q_1=2u\d_x+u_x,\quad R_3=\d_x^3.
\end{equation}
Coupling $Q_1$ and $R_3$ one obtains the Poisson pencil of the KdV hierarchy
\begin{equation}
  \Pi_{\lambda}=2u\d_x+u_x-\lambda\d_x+\epsilon^2\d_x^3
\end{equation}
discovered by Magri in \cite{magri}, while coupling $P_1$ and $R_3$ one obtains
the Poisson pencil of the Camassa--Holm hierarchy
\begin{equation}
  \tilde\Pi_{\lambda}=2u\d_x+u_x-\lambda(\d_x+\epsilon^2\d_x^3).
\end{equation}
Similarly, in the two component case one has the trio
\begin{equation}
  P_1=\begin{pmatrix}
    0 & \d_x\\ \d_x & 0 
  \end{pmatrix},\,\, Q_1=
  \begin{pmatrix}
    2u\partial_x+u_x &  v\d_x\\
    \d_x v & -2\d_x
  \end{pmatrix},\,\, R_2=\begin{pmatrix}
    0 &  -\d_x^2\\
    \d_x^2 & 0
  \end{pmatrix}
\end{equation}
Note that here the operator $R_2$ is a Dubrovin-Novikov homogeneous operator of
order two. The scheme works in the same way: one coupling yields the Poisson
pencil of the the so-called AKNS (or two-boson) hierarchy, and the other yields
the Poisson pencil of the two component Camassa-Holm hierarchy \cite{F,LZ}.
Using the language of \cite{OR} the pencils $\Pi_{\lambda}=Q_1+R_3-\lambda P_1$
and $\tilde\Pi_{\lambda}=P_1+R_3-\lambda Q_1$ are related by tri-Hamiltonian
duality. The existence of a Liouville correspondence between dual hierarchies,
generalizing the well-known transformation relating the negative flows of the
KdV hierarchy with the positive flows of the Camassa-Holm hierarchy, was
recently suggested \cite{KLOQ}.

Motivated by the above examples, in the present paper we consider the problem
of classification of compatible trios of Hamiltonian operators $P_1$, $Q_1$,
$R_{n}$ where $P_1$ and $Q_1$ are homogeneous first-order Hamiltonian operators
(also known as Hamiltonian operator of hydrodynamic type)
\begin{equation}
  \label{eq:5}
  P_1 = g^{ij}\d_x + \Gamma^{ij}_ku^k_x,\qquad
  Q_1 = h^{ij}\d_x + \Gamma^{ij}_ku^k_x,
\end{equation}
and $R_{n}$ is a homogeneous Hamiltonian operator
\begin{equation}\label{eq:10}
  R_{n} = \sum_{l=0}^{n}A^{ij}_{n,l}(u,u_x,\ldots,u_{(l)})\d_x^{(n-l)}
\end{equation}
of degree $n>1$. This means that $A^{ij}_{n,l}$ are homogeneous polynomials of
degree $l$ in the variables $u_x,\ldots,u_{(l)}$, where the homogeneous degree
is given assigning degree $1$ to the derivative w.r.t. $x$.  We recall
\cite{DN,DN2} that the homogeneity requirement implies that the operators
$P_1$, $Q_1$ and $R_n$ do not change their `form' under the action of point
transformations of the dependent variables
\begin{equation}
  \label{eq:19}
  \tilde{u}^i = \tilde{u}^i(u^j).
\end{equation}

The associated Poisson pencils are
\begin{equation}\label{eq:8}
  P_1+R_n-\lambda Q_1,\qquad P_1-\lambda (Q_1+R_n).
\end{equation}
We call a pencil of one of the above types a \emph{bi-Hamiltonian structure of
  KdV type}.  The above pencils can be thought as a deformation of a Poisson
pencil of hydrodynamic type. Due to the general theory of deformations the only
interesting cases are $n=2$ and $n=3$. In the remaining case the deformations
can be always eliminated by Miura type transformations \cite{LZ}. For this
reason we will consider only second and third order Hamiltonian operators $R_2$
and $R_3$.

We recall that second-order operators $R_2$ have been completely described in
\cite{Doyle,GP87}, and third-order operators $R_3$ have been classified in the
$m$-com\-po\-nent case with $m=1$ (in this case the operator can be reduced to
$\d_x^3$ by a point transformation~\eqref{eq:19} \cite{GP91,GP97,Doyle}) and
$m=2,3,4$ \cite{fpv1,fpv2}.

Our strategy uses the normal forms of $R_2$ and $R_3$; for each of them we will
find all possible compatible first-order Poisson pencils of hydrodynamic type
$P_1-\lambda Q_1$ and, consequently, all possible Poisson pencils of the
form~\eqref{eq:8} with $n=2$ (or $n=3$) where the three operators $P_1$, $Q_1$,
$R_2$ (or $R_3$) are mutually compatible.

In the scalar case $m=1$ there is nothing new: we obtain the KdV and
Camassa-Holm hierarchies.

In this paper we focus on the $2$-component case, leaving the $3$-component
case to future investigations. When $m=2$ there is only one homogeneous
second-order Hamiltonian operator:
\begin{equation}
  \label{eq:1}
  R_2 = \begin{pmatrix}0 & 1\\ -1 & 0\end{pmatrix}\d_x^2,
\end{equation}
and there are three homogeneous third-order Hamiltonian operators
\begin{eqnarray}
  \label{P1}
  R^{(1)}_3&=&
  \begin{pmatrix}
    0 & 1\\
    1 & 0
  \end{pmatrix}\d_x^3,\\
  \label{P2}
  R^{(2)}_3&=& \d_x
  \begin{pmatrix}
    0 & \d_x\frac{1}{u^1} \\
    \frac{1}{u^1}\d_x& \frac{u^2}{(u^1)^{2}}\d_x+\d_x\frac{u^2}{(u^1)^{2}}
  \end{pmatrix}
  \d_x,\\
  \label{P3}
  R^{(3)}_3&=& \d_x
  \begin{pmatrix}
    \d_x & \d_x \frac{u^2}{u^1}\\
    \frac{u^2}{u^1} \d_x & \frac{(u^2)^2+1}{2(u^1)^2}\d_x+
    \d_x\frac{(u^2)^2+1}{2(u^1)^{2}}
  \end{pmatrix}\d_x.
\end{eqnarray}
The operators are distinct up to transformations~\eqref{eq:19}.

Our main results are the following Theorems (the coefficients $c_i$ are
constants).
\begin{theorem}\label{th1}
  $P_1$ is a Hamiltonian operator compatible with $R_2$ if and only if
  \begin{subequations}\label{eq:12}
    \begin{align}
      &g^{11}=c_1u^1+c_2,
      \\
      &g^{12}=\f{1}{2}c_3u^1+\f{1}{2}c_1u^2+c_5
      \\
      &g^{22}=c_3u^2+c_4.
    \end{align}
  \end{subequations}
  Moreover the above metric is flat for every value of the parameters.
\end{theorem}

\begin{theorem}\label{th2}
  $P_1$ is  a Hamiltonian operator compatible with $R_3^{(1)}$ if and only if
  \begin{equation}
    \begin{split}
      g^{11} =& c_1 u^1 + c_2 u^2 + c_3,
      \\
      g^{12} =& c_4 u^1 + c_1 u^2 + c_5
      \\
      g^{22}= & c_6 u^1 + c_4 u^2 + c_7
    \end{split}\label{eq:221}
  \end{equation}
  together with the algebraic conditions specified in~\eqref{eq:165}.
\end{theorem}

\begin{theorem}\label{th3}
  $P_1$ is  a Hamiltonian operator compatible with $R_3^{(2)}$ if and only if
  \begin{subequations}\label{eq:14}
    \begin{align}
      &g^{11}=c_1 u^1 + c_2 u^2,
      \\
      &g^{12}=c_4 u^1 + \frac{c_3}{u^1} + \frac{c_2 (u^2)^2}{2 u^1},
      \\
      &g^{22}=2 c_4 u^2 + \frac{c_6}{ u^1} - \frac{c_1 (u^2)^2}{ u^1} + c_5,
    \end{align}
  \end{subequations}
  together with the algebraic conditions specified in~\eqref{eq:17}.
\end{theorem}

\begin{theorem}\label{th4}
  $P_1$ is  a Hamiltonian operator compatible with $R_3^{(3)}$ if and only if
  \begin{subequations}\label{eq:15}
    \begin{align}
      &g^{11}=c_1 u^1 + c_2 u^2+c_3,
      \\
      &g^{12}=c_4 u^1 - \frac{c_2}{2u^1} + \frac{c_3 u^2}{ u^1}+ \frac{c_2
        (u^2)^2}{ 2u^1},
      \\
      &g^{22}=2 c_4 u^2 + \frac{c_1}{u^1} + \frac{c_5 u^2}{ u^1}- \frac{c_1
        (u^2)^2}{ u^1} +c_6,
    \end{align}
  \end{subequations}
  together with the algebraic conditions specified in~\eqref{eq:18}.
\end{theorem}

The above mentioned algebraic conditions are quadratic in the parameters and
define an algebraic variety.  The problem of finding Poisson pencils of the
form \eqref{eq:8} inside the above algebraic variety is mathematically
equivalent to finding all the straight lines cointained in this variety. The
detailed list of solutions is given (case by case) in Section 3. In the generic
case we obtain:
\begin{itemize}
  \item a $5$ parameter family of mutually commuting pairs $P_1$, $Q_1$ that
    commute with $R_3^{(1)}$ (see Theorem \ref{th2d-anti} for
    further details).
  \item a $4$ parameter family of mutually commuting pairs $P_1$, $Q_1$ that
    commute with $R_3^{(2)}$ (see Theorem \ref{th3d} for
    further details).
  \item a $4$ parameter family of mutually commuting pairs $P_1$, $Q_1$ that
    commute with $R_3^{(3)}$  (see Theorem \ref{th4d} for
    further details).
\end{itemize}

The above results can also be read in the framework \cite{DZnormal} of Dubrovin
and Zhang's perturbative approach. Indeed, all the pencils that we are
considering can be regarded as deformations of a Poisson pencil of hydrodynamic
type. The classification of deformations with respect to the group of Miura
transformations
\begin{equation}\label{Miura}
  \tilde{u}^i= f^i(u^1,\ldots,u^n) +
  \sum_{k\ge1}\epsilon^k F^i_k(u,u_x,\dots,u_{(k)}),
\end{equation}
has been obtained in recent years in the semisimple case (see \cite{LZ} for the scalar case and
\cite{CPS} for the general case). It turned out that deformations are uniquely determined by their
dispersionless limit and by $n$ functions of a single variable called
\emph{central invariants}. In particular, the vanishing of the central
invariants implies the existence of a Miura transformation reducing the pencil
to its dispersionless limit. For this reason deformations with vanishing
central invariants are said to be trivial. 

In Section 4 we will first recover old and recent $2$-component examples of
bi-Hamiltonian systems of PDEs. In particular we show that the Kaup-Broer
system~\cite{K85} and a more recent multicomponent family of commuting
operators~\cite{DKT14} are particular cases of hierarchies generated by trios
with $R_2$ and that the coupled  Harry-Dym hierarchy~\cite{AF88} and the
Dispersive Water Waves system \cite{AF89} are particular cases of hierarchies
generated by trios with $R_3^{(1)}$.

Then, we provide examples of new bi-Hamiltonian systems of PDEs generated by
trios with $R_3^{(2)}$ and $R_3^{(3)}$. The systems
are expressed via rational functions; this makes them particularly interesting. 

\medskip

Computations were performed independently with Maple and with the software
package CDE \cite{cde} of the Reduce computer algebra system.

\section{Homogeneous Hamiltonian and bi-Hamil\-to\-nian
 structures}

\subsection{First-order operators and flat pencils}

First-order Hamiltonian operators of hydrodynamic type 
$$P = g^{ij}\d_x -g^{il}\Gamma^{j}_{lk }u^k_x= g^{ij}\d_x +\Gamma^{ij}_{k }u^k_x$$
have been introduced by Dubrovin and Novikov in
\cite{DN,DN2}. In the non-degenerate case ($\det(g^{ij})\neq 0$) the operator
$P$ is Hamiltonian if and only if $g_{ij}$ (the inverse of $g^{ij}$) is a flat
pseudo-Riemannian metric and $\Gamma^j_{hk}$ are the Christoffel symbols of the
 associated Levi-Civita connection.

Poisson pencils of hydrodynamic type have been introduced in the
framework of Frobenius manifolds by Boris Dubrovin in \cite{D}; they are
defined by a pair of contravariant (pseudo)-metrics $g$ and $h$ satisfying
the following conditions:
\begin{enumerate}
\item The pencil of metrics $g_{\lambda}=g-\lambda h$ is flat for any $\lambda$.
\item The (contravariant) Christoffel symbols $\Gamma^{ij}_{(\lambda) k}$ of
  the pencil $g_{\lambda}$ coincide with the pencils of Christoffel symbols:
  \begin{equation}\label{comp_christoffel}
    \Gamma^{ij}_{(\lambda) k}=\Gamma^{ij}_{(2)k}-\lambda \Gamma^{ij}_{(1)k},
  \end{equation}
  where $\Gamma^{ij}_{(1)k}$ and $\Gamma^{ij}_{(2)k}$ are the
  Christoffel symbols of the metrics $h$ and $g$ respectively.
\end{enumerate}
A pencil of contravariant metrics $g_\lambda$ fulfilling the above conditions
is called a \emph{flat pencil}. A flat pencil is said to be \emph{semisimple}
if the eigenvalues of the affinor $gh^{-1}$ are functionally independent. In
this case the eigenvalues define a special set of coordinates, called 
\emph{canonical coordinates}, where both
the metrics of the pencil become diagonal.

\subsection{Higher-order operators}

General structure theorems for higher-order homogeneous Hamiltonian
operators~\eqref{eq:10} are much weaker. We only consider the case where the
coefficient $\ell^{ij}=A^{ij}_{n,0}(u)$ of the leading term is non-degenerate:
$\det(\ell^{ij})\neq 0$. The term $A^{ij}_{n,n}(u,u_x,\ldots,u_{(n)})$ of the
above operators contains a summand of the form $d^{ij}_ku^k_{(n)}$. It can be
proved that $ - \ell_{ih}d^{hj}_k$ transform as the Christoffel symbols of a
linear connection; the fact that the operator is Hamiltonian imply that such a
connection is symmetric and flat \cite{GP91,Doyle}. In flat coordinates we have
the following canonical forms of $R_2$ and $R_3$, respectively:
\begin{equation}
  \label{eq:2}
  R_2 = \d_x \ell^{ij} \d_x,
\end{equation}
where $\ell_{ij}=T_{ijk}u^k + T^0_{ij}$ and $T_{ijk}$ are completely
skew-symmetric and
\begin{equation}
  \label{eq:11}
  R_3=\d_x\left(\ell^{ij}\d_x +c_{k}^{ij}u_{x}^{k}\right)\d_x.
\end{equation}
Moreover, introducing $c_{ijk}=\ell_{iq}\ell_{jp}c_{k}^{pq}$, the following
conditions must be fulfilled \cite{fpv1}:
\begin{subequations}\label{eq:3}
  \begin{gather}
    c_{nkm}=\frac{1}{3}(\ell_{nm,k} - \ell_{nk,m}),\label{eq:4}\\
    \ell_{mn,k} + \ell_{nk,m} + \ell_{km,n}=0,\label{eq:6}\\
    c_{mnk,l}= - \ell^{pq}c_{pml}c_{qnk} .  \label{eq:7}
  \end{gather}
\end{subequations}

Both canonical forms \eqref{eq:2} and \eqref{eq:11} are defined up to affine
transformations.  The normal forms of the operators $R_2$ and $R_3$ depend on
the number of components $m$. In the case $m=2$ we have $R_2=T_0^{ij}\d_x^2$,
where $T_0^{ij}$ is a constant skew-symmetric matrix. The operator can be
reduced to \eqref{eq:1} by an affine transformation.  There are three canonical
forms for the leading term of $R_3$ when $m=2$ modulo affine transformations
\cite{fpv1}, namely \eqref{P1}, \eqref{P2}, \eqref{P3}. One can verify that the
metric $\ell^{(2)}$ of $R_3^{(2)}$ is flat, while the metric $\ell^{(3)}$ of
$R_3^{(3)}$ is non-flat.

We stress that two homogeneous third-order Hamiltonian operators are equivalent
by a point transformation \eqref{eq:19} if and only if they have the same
normal form \eqref{P1}, \eqref{P2}, or \eqref{P3}. 
We also remark that the invariance group of $R_3$ can be enlarged to reciprocal
transformations of projective type \cite{fpv1}. When $m=2$ it can be proved
that the same projective transformation reduces the last two cases to constant
coefficients. If $m=3,4$ there is a classification of normal forms of $R_3$ up
to reciprocal transformations of projective type \cite{fpv1,fpv2}.  However,
reciprocal transformations are outside the aims of the paper.

\section{Compatible trios $P_1$, $Q_1$, $R_i$ in two components}

In this Section we classify all trios of two compatible homogeneous first-order
Hamiltonian operators $P_1$, $Q_1$ and one homogeneous Hamiltonian operator
$R_i$ of order $i$, with $i=2$ or $i=3$.

Without loss of generality we assume that the operators $R_i$ are in one of the
normal forms~\eqref{eq:1},~\eqref{P1},~\eqref{P2},~\eqref{P3}.

First of all we solve the condition $[P_1,R_i]=0$ using all coefficients
$g^{ij}$ and $\Gamma^{ij}_k$ as unknowns; it turns out that the solutions
linearly depend on a set of parameters $c_i$. Then we impose that the functions
$\Gamma^{ij}_k$ are the Chritoffel symbols of the Levi-Civita connection of
$g^{ij}$:
\begin{align}
  g^{is}\Gamma^{jk}_s = g^{js}\Gamma^{ik}_s\label{eq:9}
  \\
  \Gamma^{ij}_k + \Gamma^{ji}_k = \partial_k g^{ij}\label{eq:21}
\end{align}
In the case $i=2$ the above conditions are empty, while in the case $i=3$ we
obtain quadratic constraints for the coefficients $c_i$; in principle we should
have further restrictions coming from the flatness of $g$ but in the
two-component case this condition does not provide additional constraints (this
fact is no longer true already in the three-component case).

In order to get a compatble trio $(P_1,Q_1,R_i)$ we have to select among the pairs of flat metrics $(g,h)$
 of the above form those defining a flat pencil. Each metric is defined by a point in the space of
 parameters. We call $c_i$ the values of the parameters that provide the metric $g$  and $d_i$ the values
 of the parameters that provide the metric $h$. They can be interpreted as the coordinates of two points
 in the algebraic variety defined by the quadratic conditions described above.
  If the pair $(g,h)$ defines a flat pencil, then the straight line joining these two points is entirely
   contained in this variety. 


\begin{theorem}\label{th2d-anti}
  The Levi-Civita conditions~\eqref{eq:9},~\eqref{eq:21} for the metric
  $g^{ij}$ and the connection $\Gamma^{ij}_k$ of the operator $P_1$ that is
  compatible with $R_3^{(1)}$ are
  \begin{equation}\label{eq:165}
     c_1c_4 - c_2c_6=0,
     \quad
     c_3c_4 - c_7c_2=0,
     \quad
     c_3c_6 - c_1c_7=0.
  \end{equation}
  The above conditions imply the flatness of $g$.

  The solution of the above system is:
  \begin{enumerate}
  \item if $c_{2}\neq 0$ then $c_{6} = (c_4 c_1)/c_{2}$, $c_{7}=(c_{3}
    c_4)/c_{2}$;
  \item if $c_{2}=0$ and $c_{3}\neq 0$ then $c_{6} = (c_{7} c_1)/c_{3}$,
    $c_4=0$;
  \item if $c_{3}=0$, $c_{2}=0$ then $c_1=0$;
  \item if $c_{3}=0$, $c_{2}=0$ and $c_1\neq 0$ then $c_4=0$,
    $c_{7}= 0$.
  \end{enumerate}
The compatible pencils $g_{\lambda,kl}=g^{ij}_k - \lambda h^{ij}_l$ are
  \begin{itemize}
  \item $g_{\lambda,11}$ if $c_4=\frac{d_4 c_2}{d_2}$, or $d_3=\frac{d_2 c_3}{c_2}$, $c_1=\frac{d_1 c_2}{d_2}$;
  \item $g_{\lambda,12}$ if $d_7=\frac{d_3 c_4}{c_2}$;
  \item $g_{\lambda,13}$ if $d_6=\frac{d_4 c_1}{c_2}$, $d_7=\frac{d_4 c_3}{c_2}$.
  \item $g_{\lambda,14}$ if $d_6=\frac{c_4 d_1}{c_2}$;
  \item $g_{\lambda,22}$ if $d_7=\frac{d_3 c_7}{c_3}$, or if $d_1=\frac{d_3 c_1}{c_3}$;
  \item $g_{\lambda,23}$ if $d_4=0$, $d_6=\frac{d_7 c_1}{c_3}$;
  \item $g_{\lambda,24}$ if $d_6=\frac{c_7 d_1}{c_3}$;
  \item $g_{\lambda,33}$;
  \item $g_{\lambda,34}$ if $c_4=c_7=0$;
  \item $g_{\lambda,44}$.
  \end{itemize}  
\end{theorem}

\begin{theorem}\label{th3d}
  The Levi-Civita conditions~\eqref{eq:9},~\eqref{eq:21} for the metric
  $g^{ij}$ and the connection $\Gamma^{ij}_k$ of the operator $P_1$ that is
  compatible with $R_3^{(2)}$ are
  \begin{equation}
    c_2 c_6 + 2 c_1 c_3 = 0, \quad c_2 c_5 = 0, \quad c_1 c_5 = 0.\label{eq:17}
  \end{equation}
  The above conditions imply the flatness of $g$.

  The solution of the above system is:
  \begin{enumerate}
  \item if $c_1\neq 0$ then $c_5=0$ and $c_3=-\frac{c_2 c_6}{2 c_1}$;
  \item if $c_1=0$ and $c_2 \neq 0$ then $c_5=c_6=0$;
  \item otherwise $c_1=c_2=0$.
  \end{enumerate}
  The compatible pencils $g_{\lambda,kl}=g^{ij}_k - \lambda h^{ij}_l$ are
  \begin{itemize}
  \item $g_{\lambda,11}$ if $d_6=\frac{d_1 c_6}{c_1}$, or $d_2=\frac{d_1
      c_2}{c_1}$.
  \item $g_{\lambda,12}$ if $d_3=-\frac{d_2 c_6}{2 c_1}$.
  \item $g_{\lambda,13}$ if $d_3=-\frac{d_6 c_2}{2 c_1}$, $d_5=0$.
  \item $g_{\lambda,22}$;
  \item $g_{\lambda,23}$ if $d_5=d_6=0$.
  \item $g_{\lambda,33}$.
  \end{itemize}
\end{theorem}

\begin{theorem}\label{th4d}
  The Levi-Civita conditions~\eqref{eq:9},~\eqref{eq:21} for the metric
  $g^{ij}$ and the connection $\Gamma^{ij}_k$ of the operator $P_1$ that is
  compatible with $R_3^{(3)}$ are
  \begin{equation}
    c_2 c_5 + 2 c_1 c_3 = 0, \quad c_2 c_6 - 2 c_3 c_4 = 0,
    \quad c_1 c_6 + c_4 c_5 = 0,\label{eq:18}
  \end{equation}
  The above conditions imply the flatness of $g$.

  The solution of the above system is:
  \begin{enumerate}
  \item if $c_2\neq 0$ then $c_5=-\frac{2 c_1 c_3}{c_2}$ and $c_6=\frac{2 c_3
      c_4}{c_2}$;
  \item if $c_2=0$ and $c_3 \neq 0$ then $c_1=c_4=0$;
  \item if $c_2=c_3=0$ and $c_6 \neq 0$ then $c_1=-\frac{c_4 c_5}{c_6}$;
  \item if $c_2=c_3=c_6=0$ and $c_5 \neq0$ then $c_4=0$;
  \item otherwise $c_2=c_3=c_5=c_6=0$.
  \end{enumerate}
  The compatible pencils $g_{\lambda,kl}=g^{ij}_k - \lambda h^{ij}_l$ are
  \begin{itemize}
  \item $g_{\lambda,11}$ if $d_3=\frac{d_2 c_3}{c_2}$, or $d_1=\frac{d_2
      c_1}{c_2}$, $d_4=\frac{d_2 c_4}{c_2}$.
  \item $g_{\lambda,12}$ if $d_5=-\frac{2d_3 c_1}{ c_2}$, $d_6=\frac{2 d_3
      c_4}{c_2}$.
  \item $g_{\lambda,13}$ if $d_6=\frac{2 d_4 c_3}{2 c_2}$, with $d_4\neq
    0$, $c_3\neq 0$.
  \item $g_{\lambda,14}$ if $d_5=-\frac{2 d_4 c_3}{2 c_2}$, with $d_4\neq
    0$, $c_3\neq 0$.
  \item $g_{\lambda,15}$ if $c_3=0$.
  \item $g_{\lambda,22}$.
  \item $g_{\lambda,33}$ if $d_5=d_6=0$, or $d_5=\frac{d_6 c_5}{c_6}$, or
    $d_4=\frac{d_6 c_4}{c_6}$.
  \item $g_{\lambda,34}$ if $d_1=-\frac{d_5 c_4}{c_6}$.
  \item $g_{\lambda,35}$ if $d_1=-\frac{d_4 c_5}{c_6}$.
  \item $g_{\lambda,44}$.
  \item $g_{\lambda,45}$ if $d_4=0$.
  \item $g_{\lambda,55}$.
  \end{itemize}
  We stress that $g_{\lambda,23}$, $g_{\lambda,24}$ and
  $g_{\lambda,25}$ do not define flat pencils.
\end{theorem}

\section{Examples}
We consider some known and new examples of bi-Hamiltonian structures associated
with trios of compatibile operators. Each trio $(P_1,Q_1,R_i)$ ($i=2$,
$3$) defines two pencils $\Pi_{\lambda}=P_1+R_i-\lambda Q_1$ and
$\tilde\Pi_{\lambda}=Q_1+R_i-\lambda P_1$. In the case of new examples we compute
 the first non trivial flows of the associated
bi-Hamiltonian hierarchies.


\subsection{Case $R_2$: Cohomology spaces of curves}

In \cite{DKT14} the following six-parameter family of pairwise compatible
Hamiltonian operators defined by the cohomology spaces of curves
is considered:
\begin{equation}
  \label{eq:901}
\begin{pmatrix}
a(u^1_x + 2u^1\partial_x) + \alpha\partial_x+c\partial_x^3 &
au^2\partial_x + \beta\partial_x + \gamma\partial_x^2
\\
a\partial_x u^2 + \beta\partial_x - \gamma\partial_x^2 &
\epsilon\partial_x
\end{pmatrix}
\end{equation}
It contains
systems by Ito, Kupershmidt, Antonowicz and Fordy, Fokas and Liu, G\"umral and
Nutku.

For $\gamma=1$ and $c=0$ we have a family of commuting operators  of our type. It is easy to check that it
corresponds to the choice $c_1=2a$, $c_2=\alpha$, $c_4=\epsilon$ (and all other $c_i=0$)
 in the metric $g$ of Theorem~\ref{th1}.

\subsection{Case $R_2$: Kaup-Broer equation}

The bi-Hamiltonian property of the Kaup-Broer system was established in
\cite{K85}. The system is
\begin{equation}
  \label{eq:101}
\left\{
  \begin{matrix}
    u^1_t = ((u^1)^2/2 + u^2 + \beta u^1_x)_x,
\\
    u^2_t = (u^1u^2 + \alpha u^1_{xx} - \beta u^2_x)_x,
  \end{matrix}
\right.
\end{equation}
where $\alpha$, $\beta$ are two constants. Indeed, the system is
tri-Hamiltonian, two of the operators are of the form
\begin{equation}
  \label{eq:111}
  B_1 =
  \begin{pmatrix}
    0 & \partial_x \\ \partial_x & 0
  \end{pmatrix}
  \quad
  B_2 =
  \begin{pmatrix}
     2\partial_x & \partial_x u^1 - \partial_x^2
   \\
     u^1\partial_x + \partial_x^2 & u^2\partial_x + \partial_x u^2
  \end{pmatrix}
\end{equation}
and are defined by trio of compatible Hamiltonian operators of our class. Indeed, 
 it is easy to check that  the choice $c_2=2$, $c_3=2$ and all other $c_i$ set
to zero in the metric $g$ of Theorem~\ref{th1} yields the above example (up to
the sign of $R_2$).

According with \cite{AF89}, there exists  a Miura transformation that
brings the above system into Dispersive Water Waves  system.

\subsection{Case $R^{(1)}_3$: Dispersive Water Waves}
\label{sec:disp-water-waves}

Here we consider the example on page 482 of \cite{AF89}.  The system
\begin{align}
  \label{eq:491}
  u^1_t =& \frac{1}{4}u^2_{xxx} + \frac{1}{2}u^2u^1_x + u^1u^2_x,
  \\
  u^2_t =& u^1_x + \frac{3}{2}u^2u^2_x
\end{align}
is the DWW equation up to a Miura transformation. It is a tri-Hamiltonian
equation with respect to the operators
\begin{align}
  \label{eq:591}
  B_0 =&
  \begin{pmatrix}
    - \frac{1}{2} u^2\partial_x -\frac{1}{2}\partial_x u^2 & \partial_x
    \\
    \partial_x & 0
  \end{pmatrix}
  \\
  B_1 =&
  \begin{pmatrix}
    \frac{1}{4}\partial_x^3 + \frac{1}{2} u^1\partial_x + \frac{1}{2}\partial_x u^1
    & 0
    \\
    0 & \partial_x
  \end{pmatrix}
  \\
  B_2 = &
  \begin{pmatrix}
    0 & \frac{1}{4}\partial_x^3 + \frac{1}{2} u^1\partial_x +
    \frac{1}{2}\partial_x u^1
    \\
    \frac{1}{4}\partial_x^3 + \frac{1}{2} u^1\partial_x + \frac{1}{2}\partial_x u^1
    & \frac{1}{2}u^2\partial_x + \frac{1}{2}\partial_x u^2
  \end{pmatrix}
\end{align}
The pair $(B_0,B_2)$ is defined by a trio of compatible Hamiltonian operators of our class. Indeed, 
 if we choose $c_2=-1/2$, $c_5=1$ and all other values of $c_i$
equal to $0$ in $g$, and $d_4=1/2$ with all other values of $d_j$ equal to $0$
in $h$ we recover the above example from \eqref{eq:221}.

\subsection{Case $R^{(1)}_3$: coupled Harry-Dym hierarchy}

We consider the example on page L273 of \cite{AF88}. The system
\begin{align}
 u^1_1 =& \left(\frac{1}{4(u^2)^{1/2}}\right)_{xxx} -
  \alpha\left(\frac{1}{(u^2)^{1/2}}\right)_x\label{eq:191}
  \\
  u^2_t = & u^1 \left(\frac{1}{(u^2)^{1/2}}\right)_x + \frac{u^{1}_x}{2(u^2)^{1/2}}
\end{align}
is tri-Hamiltonian with respect to the following operators
\begin{align}
  \label{eq:291}
  B_0= &
  \begin{pmatrix}
    -\frac{1}{2}u^1\partial_x -\frac{1}{2}\partial_x u^1 &
    -\frac{1}{2}u^2\partial_x -\frac{1}{2}\partial_x u^2
    \\
    -\frac{1}{2}u^2\partial_x -\frac{1}{2}\partial_x u^2 & 0
  \end{pmatrix},
  \\
  \label{eq:391}
  B_1 = &
  \begin{pmatrix}
    \frac{1}{4}\partial_x^3 - \alpha\partial_x & 0
    \\
    0 & -\frac{1}{2}u^2\partial_x -\frac{1}{2}\partial_x u^2
  \end{pmatrix}
  \\
  B_2 = &
  \begin{pmatrix}
    0 & \frac{1}{4}\partial_x^3 - \alpha\partial_x
    \\
    \frac{1}{4}\partial_x^3 - \alpha\partial_x & +\frac{1}{2}u^1\partial_x
    +\frac{1}{2}\partial_x u^1
  \end{pmatrix}
\end{align}
The pair $(B_0,B_2)$ is defined by a trio of compatible Hamiltonian operators of our class. Indeed, 
if we choose $c_1=-1/2$ with all other $c_i$ equal to $0$ in $g$ and
$d_5=-\alpha$, $d_6=1/2$ with all other $d_j$ equal to $0$ in $h$ we recover
the above example from \eqref{eq:221}.

\subsection{Case $R^{(2)}_3$: pencil $g_{\lambda,11}$}
\label{sec:case-r2_3:-pencil}

Choosing
$$c_4=0,\quad c_1=-1,\qquad c_6=-1,\qquad c_2=0,\qquad d_2=0,\qquad d_1=0$$
we obtain the trio
\begin{eqnarray*}
  P_1&=&
  \begin{pmatrix} 
    - u^{1} & 0 \\
    0 & \frac{(u^{2})^2 - 1}{u^1}
  \end{pmatrix}\d_x+
  \frac{1}{2}\begin{pmatrix}-u^1_x
    &
    u^2_x\\
    -u^2_x &
    \frac{2  u^{1}  u^{2}  u^2_x - (u^{2})^{2}u^1_x + u^1_x}{(u^{1})^{2}}
  \end{pmatrix}
  \\
  Q_1&=&
  \begin{pmatrix}
    0  & - u^1  \\
    - u^1 & - 2u^2
  \end{pmatrix}\d_x +
  \begin{pmatrix}
    0 & - u^1_x\\
    0 & - u^2_x
  \end{pmatrix}
  \\
  R^{(3)}_3&=& \d_x
  \begin{pmatrix}
    0 & \d_x\frac{1}{u^1} \\
    \frac{1}{u^1}\d_x& \frac{u^2}{(u^1)^{2}}\d_x+\d_x\frac{u^2}{(u^1)^{2}}
  \end{pmatrix}
  \d_x.
\end{eqnarray*}
Starting from the Casimirs of $Q_1$
\begin{displaymath}
  C_1=\int_{S^1}u^1\,dx,
  \qquad
  C_2=\int_{S^1}\f{u^2}{u^1}\,dx,
\end{displaymath}
the first flows of the bi-Hamiltonian hierarchy are
$$u_{t_i}=(P_1+\epsilon^2 R_3)\delta C_i,\qquad i=1,2,$$
that is
\begin{eqnarray*}
  u^1_{t_1}=-\f{1}{2}u^1_x,\qquad
  u^2_{t_1}=-\f{1}{2}u^2_x
\end{eqnarray*}
and
\begin{eqnarray*}
  u^1_{t_2}&=&\f{3}{2}\f{u^2_x}{u^1}-\f{3}{2}\f{u^2u^1_x}{(u^1)^2}-\f{u^1_{xxx}}{(u^1)^3}+9\f{u^1_xu^1_{xx}}{(u^1)^4}-12\f{(u^1_x)^3}{(u^1)^5}\\
  u^2_{t_2}&=&\f{3}{2}\f{(1-(u^2)^2)u^1_x}{(u^1)^3}+\f{3}{2}\f{u^2u^2_x}{(u^1)^2}-\f{30u^2(u^1_{x})^3}{(u^1)^6}+10\f{u^2_x(u^1_{x})^2}{(u^1)^5}+12\f{u^2_x(u^1)_x^2}{(u^1)^5}+\\
  &&-\f{3u^2_{x}u^1_{xx}}{(u^1)^4}-2\f{u^2u^1_{xxx}}{(u^1)^4}-\f{u^2_{xx}u_x^1}{(u^1)^4}.
\end{eqnarray*}

\subsection{Case $R^{(2)}_3$: pencil $g_{\lambda,13}$}
\label{sec:case-r2_3:-pencil-1}

Choosing
$$c_3=0,\quad d_3=1,\quad c_2=2,\quad c_4=1,\quad d_4=0,\quad d_5=0$$
we obtain the trio
\begin{eqnarray*}
  P_1&=&
  \begin{pmatrix} 
    2 u^{2} & \frac{(u^{1})^2 + (u^{2})^2}{u^1} \\
    \frac{(u^{1})^2 + (u^{2})^2}{u^1} & 2u^{2}
  \end{pmatrix}\d_x+
  \begin{pmatrix}
    u^2_x & u^1_x \\
    \frac{u^2(2 u^1 u^2_x - u^1_x u^2)}{(u^1)^2} & u^2_x
  \end{pmatrix}
  \\
  Q_1&=&
  \begin{pmatrix} 
    0  &  -1/u^1  \\
    - 1/u^1 & 0
  \end{pmatrix}\d_x+
  \begin{pmatrix}
    0 & 0 \\ \frac{u^1_x}{(u^1)^2} & 0
  \end{pmatrix}
  \\
  R^{(2)}_3&=& \d_x
  \begin{pmatrix}
    0 & \d_x\frac{1}{u^1} \\
    \frac{1}{u^1}\d_x& \frac{u^2}{(u^1)^{2}}\d_x+\d_x\frac{u^2}{(u^1)^{2}}
  \end{pmatrix}
  \d_x.
\end{eqnarray*}
Starting from the Casimirs of $Q_1$
\begin{displaymath}
  C_1=\int_{S^1}\f{1}{2}(u^1)^2\,dx,
  \qquad
  C_2=\int_{S^1}u^2\,dx,
\end{displaymath}
the first flows of the bi-Hamiltonian hierarchy are
\begin{eqnarray*}
  u^1_{t_1}&=&u^1_x\\
  u^2_{t_1}&=&u^2_x
\end{eqnarray*}
and
\begin{eqnarray*}
  u^1_{t_2}&=&2u^2u^1_x+u^1u^2_x\\
  u^2_{t_2}&=&u^1u^1_x+2u^2u^2_x-\f{u^1_xu^1_{xx}}{(u^1)^2}+\f{u^1_{xxx}}{u^1},
\end{eqnarray*}
respectively.

\subsection{Case $R^{(3)}_3$: pencil $g_{\lambda,12}$}
\label{sec:case-r3_3:-pencil}

Choosing
$$c_1=1,\quad c_2=-1,\quad d_3=1,\quad c_3=0,\quad c_4=0$$
we obtain the trio
\begin{align*}
\begin{split}
  P_1=&
  \begin{pmatrix} 
    u^{1} - u^{2} & \frac{- (u^{2})^2+1}{2u^1} \\
    \frac{- (u^{2})^2+1}{2u^1} & \frac{- (u^{2})^2+1}{u^1}
  \end{pmatrix}\d_x+
\\
  &\frac{1}{2}\begin{pmatrix}
    u^1_x - u^2_x  & - u^2_x \\ 
    \frac{(u^{1})^{2}  u^2_x
      - 2 u^{1} u^{2} u^2_x + u^1_x  (u^{2})^{2} - u^1_x}{(u^{1})^{2}}&
    \frac{-2u^{1}u^{2}u^2_x + u^1_x(u^{2})^{2} - u^1_x}{(u^{1})^{2}}
  \end{pmatrix}
\end{split}
  \\
  Q_1=&
  \begin{pmatrix}
    - 1  &  - \frac{u^2}{u^1}  \\
    - \frac{u^2}{u^1} & - 2\frac{u^2}{u^1}
  \end{pmatrix}\d_x+
  \begin{pmatrix}
    0&0\\ \frac{ - u^1u^2_x + u^1_xu^2}{(u^1)^2}& \frac{ - u^1u^2_x + u^1_x
      u^2}{(u^1)^{2}}
  \end{pmatrix}
  \\
  R_3^{(3)}=& \d_x
  \begin{pmatrix}
    1 & \d_x \frac{u^2}{u^1}\\
    \frac{u^2}{u^1} \d_x & \frac{(u^2)^2+1}{2(u^1)^2}\d_x+
    \d_x\frac{(u^2)^2+1}{2(u^1)^{2}}
  \end{pmatrix}\d_x.
\end{align*}
Starting from the Casimirs of $Q_1$
\begin{displaymath}
  C_1=\int_{S^1}(u^1-u^2)\,dx,
  \qquad
  C_2=\int_{S^1}\sqrt{(u^2)^2-2u^1u^2}\,dx,
\end{displaymath}
one easily gets the first non trivial flows of the associated bi-Hamiltonian
hierarchy.

\section{Appendix: central invariants}
Let
\begin{equation}\label{pencil}
  \begin{array}{l}
    \displaystyle
    \Pi^{ij}_{\lambda}=\omega^{ij}_{\lambda}+\sum_{k\ge 1}\epsilon^k\sum_{l=0}^{k+1}A^{ij}_{2;k,l}(u,u_x,\dots,u_{(l)})\d_x^{(k-l+1)}\\
    \displaystyle
    \qquad
    -\lambda\sum_{k\ge 1}\epsilon^k\sum_{l=0}^{k+1}A^{ij}_{1;k,l}(u,u_x,\dots,u_{(l)})\d_x^{(k-l+1)},
  \end{array}
\end{equation}
($A^{ij}_{1;k,l}$ and $A^{ij}_{2;k,l}$ are homogeneous differential polynomials
of degree $l$) be a deformation of a semisimple Poisson pencil of hydrodynamic
type 
$$\omega^{ij}_{\lambda}=(g_2^{ij}-\lambda g_1^{ij})\d_x+(\Gamma^{ij}_{(2)k}-\lambda\Gamma^{ij}_{(1)k})u^k_x.$$.

The central invariants are then defined as \cite{LZ}:
$$
s_i=\frac{1}{(f^i)^2} \left(A^{ii}_{2;2,0}-r^i A^{ii}_{1;2,0}+\sum_{k\neq i}
  \frac{(A^{ki}_{2;1,0}-r^i A^{ki}_{1;1,0})^2}{f^k (r^k-r^i)} \right),
 \quad i=1,\ldots,n,
$$
where $f^i$ are the diagonal components of the contravariant metric $g_1$ in
canonical coordinates.

The main result of \cite{LZ} is the following: {\itshape Two deformations of the
  same Poisson pencil of hydrodynamic type are related by a Miura
  transformation if and only if their central invariants coincide. In
  particular deformations $\Pi_{\lambda}$ with vanishing central invariant can
  be reduced to their dispersionless limit $\omega_{\lambda}$ by a Miura
  transformation. This means that there exists a transformation of the form
  \begin{equation}\label{Miura2}
    \tilde{u}^i= u^i+\sum_{k\ge1}\epsilon^k
  F^i_k(u,u_x,\dots,u_{(k)}),
  \end{equation}
  (where $F^i_k(u,u_x,\dots,u_{(k)})$ are homogeneous differential polynomials
  of degree $k$) such that
$$
\Pi_{\lambda}^{ij}=L^{*i}_k\omega_{\lambda}^{kl}L^j_l,
$$
where
$$L^i_k=\sum_s(-\d_x)^s\f{\d\tilde{u}^i}{\d u^{(k,s)}},\qquad
L^{*i}_k=\sum_s\f{\d\tilde{u}^i}{\d u^{(k,s)}}\d_x^s.$$.}

Let us now apply the above result to the new examples obtained in the previous
Section (Subsections \ref{sec:case-r2_3:-pencil},
\ref{sec:case-r2_3:-pencil-1}, \ref{sec:case-r3_3:-pencil}).

In the first example the canonical coordinates are
\begin{eqnarray*}
  \lambda^1=\frac{u^2+1}{u^1},\qquad\lambda^2=\frac{u^2-1}{u^1}
\end{eqnarray*}
and the central invariants are
\begin{displaymath}
  s_1 = \f{1}{2},\qquad s_2 = - \f{1}{2}.
\end{displaymath}

In the second example the canonical coordinates are
\begin{displaymath}
  \lambda^1 = (u^1+u^2)^2,
  \qquad
  \lambda^2 = (u^1-u^2)^2,
\end{displaymath}
and the central invariants are
\begin{displaymath}
  s_1 = -\frac{1}{8\sqrt{\lambda^1}},
  \qquad
  s_2 = \frac{1}{8\sqrt{\lambda^2}}.
\end{displaymath}

In the last example the canonical coordinates are
\begin{displaymath}
  \lambda^1 = -\frac{1}{2}\,{\frac {(u^2)^{2}-1}{u^2}},
  \qquad
  \lambda^2 = \frac{1}{2} \frac{4\,(u^1)^{2}-4\,u^1u^2 + (u^2)^2-1}{2u^1-u^2},
\end{displaymath}
and the central invariants are
\begin{displaymath}
  s_1 = \frac{1}{2}\frac{\lambda^1\sqrt {(\lambda^1)^{2}+1}-(\lambda^1)^{2}-1}
  {(\lambda^1)^{2}+1},
  \qquad
  s_2 = - \frac{1}{2}
  \frac {\lambda^2\sqrt{(\lambda^2)^{2}+1} + (\lambda^2)^{2}+1}
  {(\lambda^2)^{2}+1}.
\end{displaymath}

This means that all the new examples of Poisson pencils obtained in the previous Section are not Miura-trivial.

\section*{Acknowledgements}

We thank Alberto Della Vedova for useful discussions.  PL and RV acknowledge
financial support from GNFM. RV acknowledge also financial support from INFN by
IS-CSN4 \emph{Mathematical Methods of Nonlinear Physics} and from Dipartimento
di Matematica e Fisica ``E. De Giorgi'' of the Universit\`a del Salento.

\end{document}